\documentclass[fleqn,10pt]{article} 

\usepackage{graphicx} 
\usepackage{dcolumn}  
\usepackage{bm}       
\usepackage{amsmath}
\usepackage[latin1]{inputenc}
\usepackage[T1]{fontenc}
\usepackage{cite} 

\begin{document}

\begin{center}
\LARGE
\textbf{Positron spectroscopy of point defects in the skyrmion-lattice compound MnSi}\\

\vspace{0.5cm}

\small
Markus Reiner$^{1,2}$, Andreas Bauer$^1$, Michael Leitner$^2$, Thomas Gigl$^{1,2}$, \\
Wolfgang Anwand$^3$,Maik Butterling$^3$, Andreas Wagner$^3$, Petra Kudejova$^2$, \\Christian Pfleiderer$^1$, and
Christoph Hugenschmidt$^{1,2}$\footnote{E-mail: christoph.hugenschmidt@frm2.tum.de}

\vspace{0.5cm}
$^1$Physik-Department, Technische Universit\"at M\"unchen, 85748 Garching, Germany\\
$^2$Heinz Maier-Leibnitz Zentrum (MLZ), Technische Universit\"at M\"unchen, \\85748 Garching, Germany\\
$^3$Institut f\"ur Strahlenphysik, Helmholtz-Zentrum Dresden-Rossendorf, \\01314 Dresden, Germany\\
\end{center}
\date{\today}
\vspace{0.3cm}



\begin{abstract}
Outstanding crystalline perfection is a key requirement for the formation of new forms of electronic order in a vast number of widely different materials. Whereas excellent sample quality represents a standard claim in the literature, there are, quite generally, no reliable microscopic probes to establish the nature and concentration of lattice defects such as voids, dislocations and different species of point defects on the level relevant to the length and energy scales inherent to these new forms of order. Here we report an experimental study of the archetypical skyrmion-lattice compound MnSi, where we relate the characteristic types of point defects and their concentration to the magnetic properties by combining different types of positron spectroscopy with ab-initio calculations and bulk measurements. We find that Mn antisite disorder broadens the magnetic phase transitions and lowers their critical temperatures, whereas the skyrmion lattice phase forms for all samples studied underlining the robustness of this topologically non-trivial state. Taken together, this demonstrates the unprecedented sensitivity of positron spectroscopy in studies of new forms of electronic order.
\end{abstract}

\flushbottom
%
%
\thispagestyle{empty}

\section*{Introduction}
\normalsize
Intermetallic compounds attract tremendous interest as a playground for novel electronic phases, such as unconventional superconductivity, partial forms of the electronic and spin order, emergent behaviour such as magnetic monopoles in the spin ices and frustration in  spin liquids.
In general, the preparation of high-quality single crystals of the materials of interest is extremely demanding, since a large number of mechanisms lead to the formation of intrinsic point defects. Examples include the formation due to configuration entropy,  kinetically inhibited atomic motion during solidification and frozen-in disorder.
The different species of defects comprise of, e.g., point defects such as vacancies and antisite disorder, dislocations or vacancy clusters.
Despite their great relevance, the qualitative and quantitative identification of point defects has been very challenging technically. 
For instance, scanning tunneling microscopy is a surface sensitive probe only, whereas transmission electron microscopy introduces ambiguities, as the sample has to be machined thinned and prepared mechanically.  
Further, the degree of long-range order is conventionally accessible by diffraction techniques. Yet, point defect concentrations cannot be resolved below 10$^{-2}$.

In this paper, we demonstrate how to obtain microscopic insights on the existence, nature and quantity of defects in intermetallic compounds.
We combine various positron-based techniques and ab-initio calculations of point defect thermodynamics for the identification of the defect species and the quantitative determination of the defect density.
This allows us to elucidate their influence on the magnetic and transport properties in the itinerant-electron magnet MnSi.

For our study we have selected MnSi as an important show-case of several aspects of new forms of electronic order. Notably, MnSi and related compounds in recent years have attracted great interest due to the formation of a regular arrangement of spin whirls forming a so-called skyrmion lattice at finite magnetic fields in a small phase pocket just below $T_{c}$~\cite{2009:Muhlbauer:Science, 2010:Munzer:PhysRevB, 2010:Pfleiderer:JPhysCondensMatter, 2010:Yu:Nature, 2011:Yu:NatureMater, 2012:Seki:Science, 2012:Adams:PhysRevLett, 2013:Nagaosa:NatureNano, 2015:Schwarze:NatureMater, 2015:Tokunaga:arXiv}. The non-trivial topological winding of these whirls gives rise to an emergent electrodynamics, in which each skyrmion carries one quantum of emergent magnetic flux~\cite{2012:Schulz:NaturePhys, 2010:Jonietz:Science, 2012:Yu:NatCommun}. The emergent electrodynamics is at the heart of exceptionally strong spin transfer torques overcoming defect related pinning. This has generated great interest to exploit skyrmions in spintronics devices.
Another important facet of the magnetic properties of MnSi is the paramagnetic to helimagnetic transition, which represents the first unambiguous example of a fluctuation-induced first order transition as long predicted by Brazovskii in the context of soft matter. As for the skyrmion lattice phase the effects of tiny defect concentrations appear to be of central importance. 
Last but not least, high pressure studies in MnSi have revealed an extended non-Fermi liquid regime, where both the exponent and prefactor of the resistivity are essentially independent of the residual resistivity~\cite{2001:Pfleiderer:Nature, 2003:Doiron-Leyraud:Nature, 2007:Pfleiderer:JLowTempPhys}. 
These various aspects of the importance of defects are underscored by the metallurgical properties of MnSi, which appear to be particularly amenable to the preparation of large, virtually perfect single crystals. In turn, MnSi represents a show-case par excellence in which the influence of low defect concentrations on novel electronic properties may be studied. 

MnSi belongs to the cubic B20 compounds with the space group $P2_{1}3$ lacking inversion symmetry.
MnSi	melts congruently at $1270\,^{\circ}\mathrm{C}$ and large single crystals may be readily grown from the melt using methods such as Czochralsky, Bridgman, and inductively or optically heated float-zoning~\cite{1991:Okamoto:JPhaseEquilib}. 
The resulting samples readily show residual resistivity ratios (RRR) between 50 and 100, where carefully annealed specimens reach values as high as 1000~\cite{2003:Doiron-Leyraud:Nature}. 
At ambient pressure, however, small deviations of the transition temperatures and the shapes of the corresponding thermodynamic signatures were reported for samples with different RRR~\cite{2001:Pfleiderer:JMagnMagnMater, 2011:Stishov:PhysUsp, 2012:Bauer:PhysRevB}. The latter also seems to influence the possibility to metastably extend the skyrmion lattice pocket at large hydrostatic pressures under field-cooling~\cite{2013:Ritz:PhysRevB}. However, neither the density nor the type of lattice defects in single-crystal MnSi has been addressed before. Knowledge of the defect structure in turn is key to understanding phenomena such as the very weak energy scales controlling the Brazovskii transition, the depinning of the skyrmion lattice under electrical currents~\cite{2013:Iwasaki:NatCommun, 2013:Lin:PhysRevB, 2015:Muller:PhysRevB}, the dynamics of the topological unwinding of skyrmions~\cite{2013:Milde:Science}, or the emergence of non-Fermi liquid behavior at high pressures~\cite{2013:Ritz:Nature}.

Given the importance of any of these topics, we decided to study the point defects in MnSi in a series of optically float-zoned single crystals by means of positron annihilation spectroscopy. In combination with calculations of the effective formation energies of point defects this allows us to determine the species and density of defects quantitatively. Measurements of the specific heat and the ac susceptibility demonstrate that the RRR is not sufficient to characterize a sample of MnSi, as specimens with the same RRR but different dominant defect species show discrepancies in the thermodynamic signatures of the magnetic phase transitions.

\section*{Results and Discussion}

\begin{figure}[b!]
\includegraphics[width=0.85\textwidth]{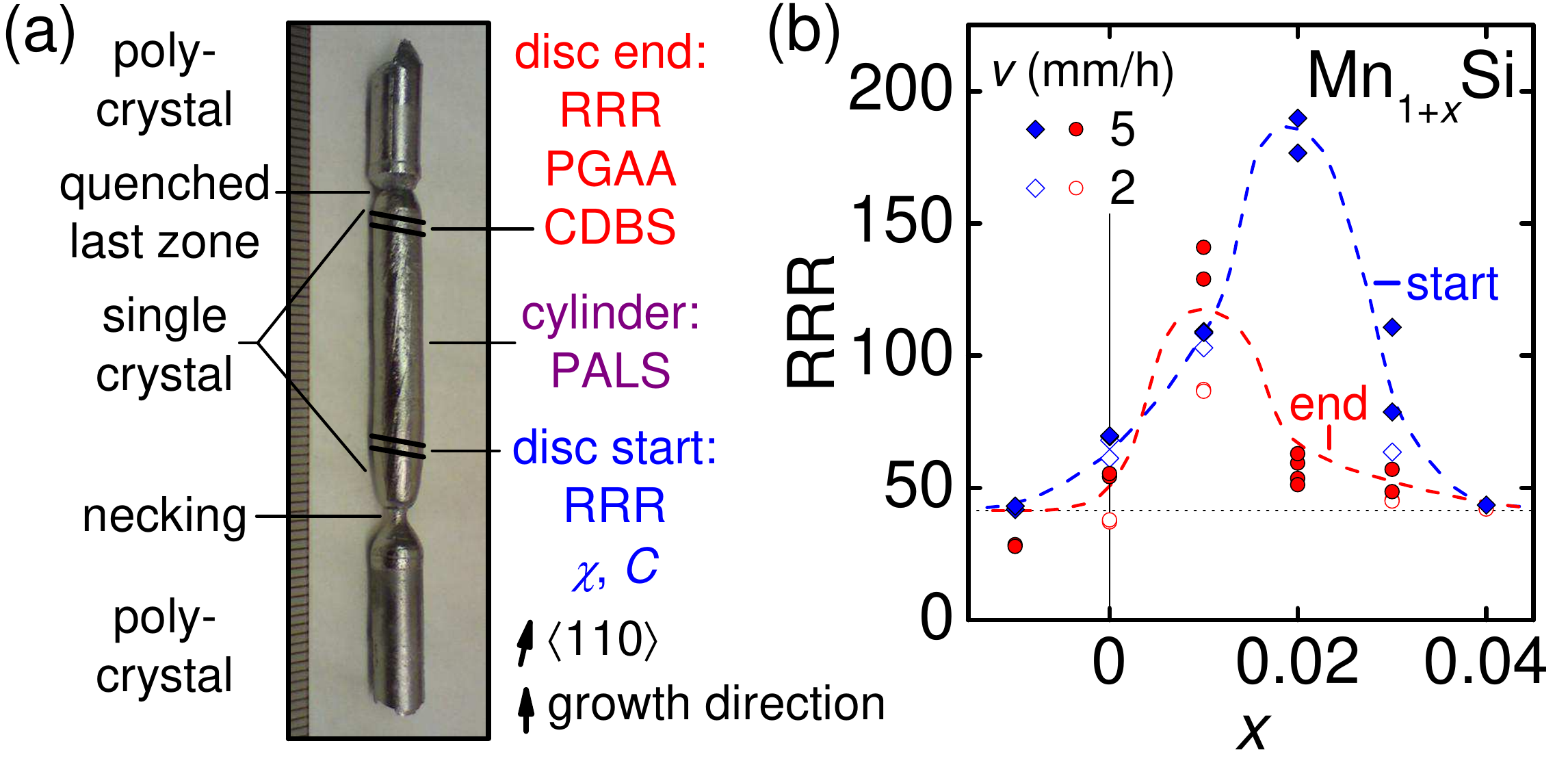}
\caption{(Color online) Preparation of the MnSi specimens. (a)~Photograph of a float-zoned ingot. Single-crystal discs from the start and the end as well as the remaining cylinders were studied. (b)~RRR as function of the initial Mn excess $x$ for different growth rates $v$. Samples from the same disc exhibit very similar RRRs. Dashed lines are guides to the eye. Here and in the following figures, data acquired on discs from start, end, and the cylinders are represented
respectively.\label{figure1}}
\end{figure}

For our study ten single crystals of MnSi were grown by means of optical float-zoning~\cite{2011:Neubauer:RevSciInstrum, 2014:Bauer:PhD}. We investigated the influence of two growth parameters, namely the growth rate, $v$, and the composition of the feed rods represented by the Mn excess, $x$, of the initial net weight of Mn$_{1+x}$Si, $-0.01 \leq x \leq 0.04$. We induced this Mn excess in order to compensate evaporative losses of Mn caused by its high vapor pressure during crystal preparation. The cylinders of the single crystals (see Fig.~\ref{figure1}(a)) were investigated by positron annihilation lifetime spectroscopy (PALS) and the end discs by coincident Doppler broadening spectroscopy (CDBS). Prompt gamma activation analysis (PGAA)~\cite{2015:Revay:NuclInstrumMethA} results were consistent with a stoichiometric composition within the estimated error of 1.5\,at.\%. For all specimens we expect the same total density of impurity atoms originating from the starting elements below $5\cdot10^{-5}$.

The RRR of all samples studied is shown in Fig.~\ref{figure1}(b). Additionally, the values for all crystals are given in Tab.~\ref{tab:values} together with the other important measured parameters introduced later. The start (blue) and end (red) of the single crystals exhibit qualitatively similar behavior, where both low and high values of $x$ lead to low RRRs around 40. A maximum in the RRR, expected for a minimal total defect concentration, is obtained for a slight initial Mn excess as consistent with a compensation of the observed evaporation of Mn during crystal growth. The small systematic variations between the start and the end of the crystals may be attributed to the different temperature history during growth where the start stayed at elevated temperatures for a longer period of time leading to larger losses of Mn. In the parameter range studied, the growth rate had no significant influence on the RRR.

\begin{figure}[b!]
\includegraphics[width=0.85\textwidth]{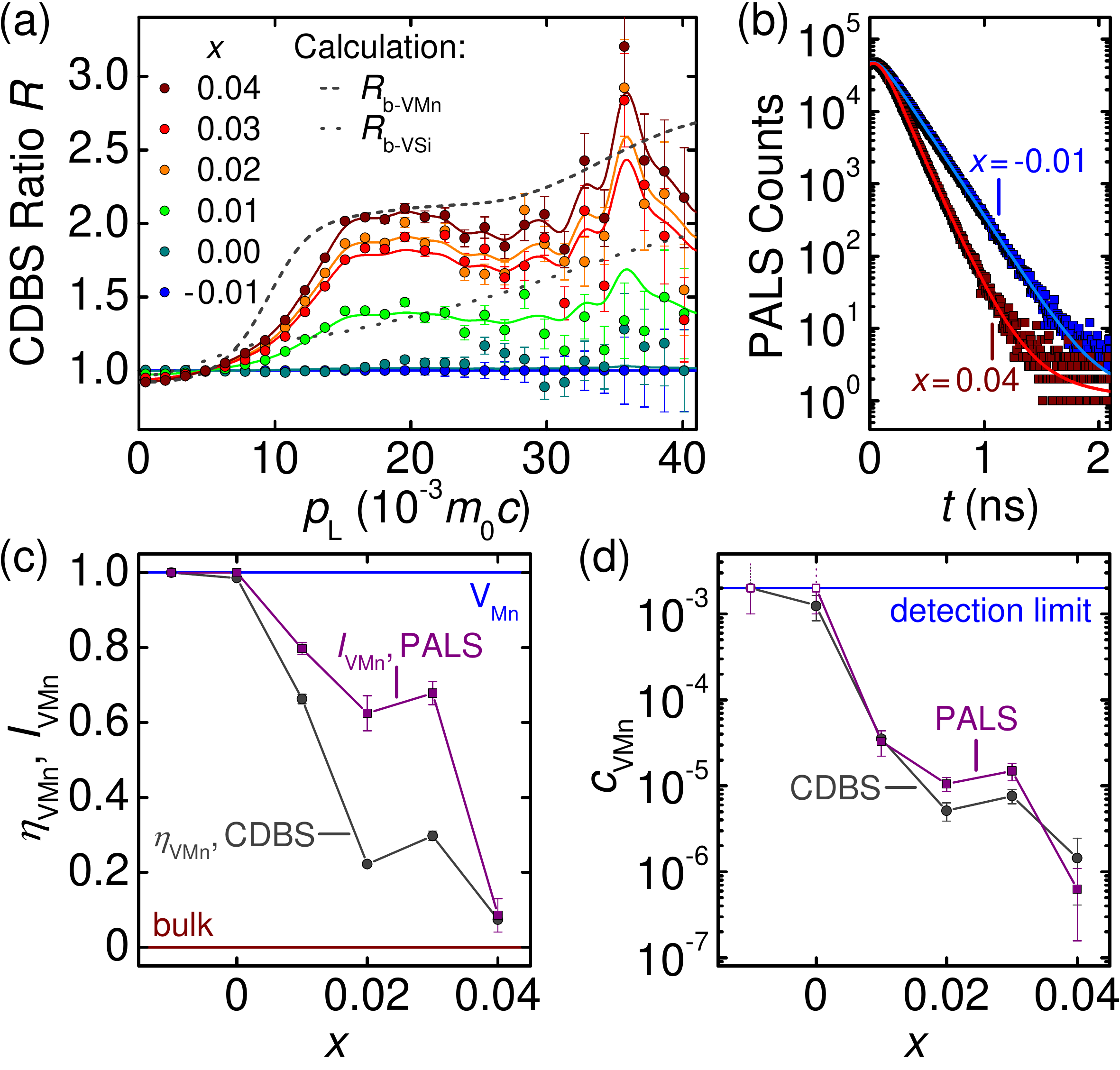}
\caption{(Color online) Results of the positron annihilation spectroscopy. (a)~CDBS ratio curves $R_{\mathrm{x}}$ as function of the longitudinal momentum component $p_{\mathrm{L}}$. Solid lines represent linear superpositions of $R_{-0.01}$ and $R_{0.04}$, dashed lines the calculated curves $R_{\text{b-VMn}}$ and $R_{\text{b-VSi}}$. (b)~PALS spectra together with fits to the data (solid lines). (c)~Fraction $\eta_{\mathrm{VMn}}$ and PALS intensity $I _{\mathrm{VMn}}$ of positrons annihilating in Mn vacancies, V$_{\mathrm{Mn}}$, as function of $x$. (d) Concentration $c_{\mathrm{VMn}}$ of V$_{\mathrm{Mn}}$ obtained from CDBS and PALS.\label{figure2}}
\end{figure}

The detected normalized CDB spectra are shown in Fig.~\ref{figure2}(a) after division by the spectrum for $x = -0.01$ as so-called ratio curves, $R_{x}$. The spectra are described very well by linear combinations of $R_{-0.01}$ and $R_{0.04}$ as depicted by solid lines. Here, the fitted weighting factor $f_{x}$ is defined by $f_{-0.01} = 0$ and $f_{0.04} = 1$. The measured ratio curves $R_{x}$ were compared to calculated ones, namely $R_{\text{b-VMn}}$ and $R_{\text{b-VSi}}$ of defect free bulk MnSi using either the spectra characteristic for annihilation in Mn vacancies $V_{\mathrm{Mn}}$ or Si vacancies $V_{\mathrm{Si}}$ as reference. As shown in Fig.~\ref{figure2}(a), $R_{\text{b-VMn}}$ describes well $R_{0.04}$ for large momenta, $p_{\mathrm{L}} > 15\cdot10^{-3}\,m_{0}c$, where the calculation method works with high reliability~\cite{2002:Tang:PhysRevB, 2006:Makkonen:PhysicaB, 2006:Makkonen:PhysRevB, 2014:Reiner:JPhysConfSer}. Hence, for $x = 0.04$ most positrons annihilate in the bulk. For $x \leq 0$ the spectra hardly differ, suggesting that essentially all positrons annihilate in V$_{\mathrm{Mn}}$ (saturation trapping).

In order to determine the fraction, $\eta_{\mathrm{b}}$, of positrons annihilating in the defect-free bulk for $x=0.04$, we compared the areas enveloped by $R_{\text{b-VMn}}-1$ and $R_{0.04}-1$ in the range $15.9\cdot10^{-3}\,m_{0}c < p_{\mathrm{L}} < 39.5\cdot10^{-3}\,m_{0}c$. We deduce that in this sample $93(5)\,\%$ of the positrons annihilate in defect-free MnSi and hence $\eta _{\mathrm{b}}(x) = 0.93 \cdot f_{x}$ for the other samples. Consequently, the residual fraction $\eta_{\mathrm{VMn}}(x) = 1-\eta_{\mathrm{b}}(x)$ depicted in Fig.~\ref{figure2}(c) is attributed to positrons annihilating in V$_{\mathrm{Mn}}$. We finally determined the concentration $c_{\mathrm{VMn}}(x) = (\eta_{\mathrm{b}}(x)^{-1} - 1) / (\tau_{\mathrm{b}}\mu_{\mathrm{VMn}})$, shown in Fig.~\ref{figure2}(d), by adapting the commonly used trapping model~\cite{1995:Hautojarvi:Book}. Here, we used the calculated bulk lifetime $\tau_{\mathrm{b}}=111$\,ps and an assumed trapping coefficient $\mu_{\mathrm{VMn}}=5\cdot10^{14}\,\mathrm{s}^{-1}$ (Typical values of $\mu$ are between $10^{14}\,\mathrm{s}^{-1}$ and $10^{15}\,\mathrm{s}^{-1}$~\cite{1995:Hautojarvi:Book}).

PALS spectra, exemplary shown in Fig.~\ref{figure2}(b), display the same trend. For $x \leq 0$ only one lifetime component is found with $\tau_{\mathrm{VMn}} = 185(4)$\,ps in excellent agreement with the calculated value of 181\,ps for a positron trapped in V$_{\mathrm{Mn}}$. For $x = 0.04$, $\tau_{\mathrm{VMn}}$ contributes only with an intensity of 8.5\,\% to the lifetime spectrum and the dominant extracted lifetime $\tau_{\mathrm{b}} = 119(3)$\,ps is close to the calculated bulk value of 111\,ps. Both lifetimes, $\tau_{\mathrm{VMn}}$ and $\tau_{\mathrm{b}}$, were fixed for analyzing the remaining crystals. Here, we determined the intensity, $I_{\mathrm{VMn}}(x)$, shown in Fig.~\ref{figure2}(c) arising from annihilation in V$_{\mathrm{Mn}}$ for the different samples using the fitted positron trapping rate $\kappa_{\mathrm{VMn}}(x)$ from bulk into V$_{\mathrm{Mn}}$. As in CDBS, for $x > 0$ annihilation in both defect-free bulk and V$_{\mathrm{Mn}}$ is observed, which allowed us to calculate the concentration $c_{\mathrm{VMn}}(x) = \kappa_{\mathrm{VMn}}(x) / \mu_{\mathrm{VMn}}$, shown in Fig.~\ref{figure2}(d).

The values of $c_{\mathrm{VMn}}$ obtained in cylinders by PALS and discs by CDBS agree very well implying a homogeneous distribution of V$_{\mathrm{Mn}}$ along the cylindrical samples. For $x \leq 0$ only a lower limit of $2\cdot10^{-3}$ may be given for $c_{\mathrm{VMn}}$ due to saturation trapping of positrons in V$_{\mathrm{Mn}}$. For $x = 0.04$, the minimal vacancy concentration of $c_{\mathrm{VMn}} \approx 1\cdot10^{-6}$ establishes that an initial Mn excess efficiently suppresses the formation of V$_{\mathrm{Mn}}$ during crystal growth.

\begin{figure}
\includegraphics[width=0.9\textwidth]{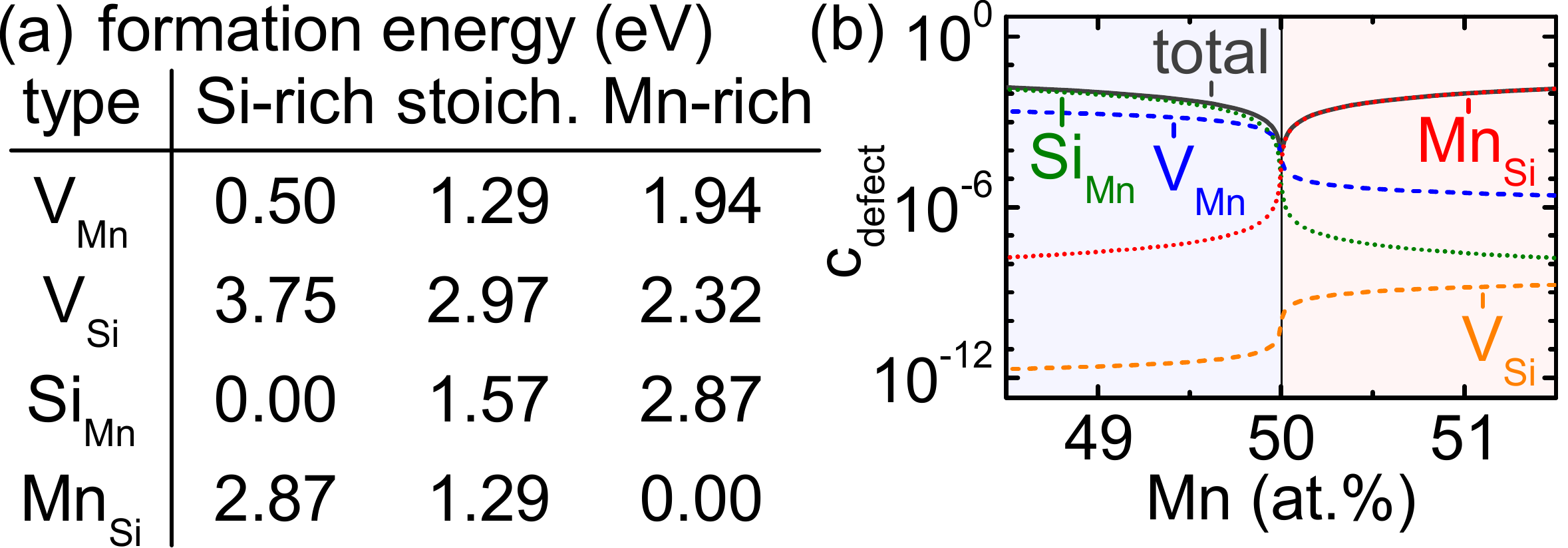}
\caption{(Color online) Calculations of point defect energetics. (a)~Effective formation energies in MnSi with different composition. (b)~Expected defect concentrations as a function of composition for MnSi around its melting temperature. \label{figure3}}
\end{figure}

In the appropriate grand-canonical theory \cite{1999:Meyer:PhysRevB}, the internal energies of the relaxed point defect configurations yield effective formation energies for the respective defect species. Their calculated values, as summarized in Fig.~\ref{figure3}(a), show that MnSi belongs to the class of compounds with antisite accommodation of deviations from stoichiometry (with the defect energetics described by the two dimensionless parameters $\xi=0.495$ and $\eta=-0.318$)~\cite{2015:Leitner:arXiv}. 
Specifically, also on the Si-rich side deviations from stoichiometry are accommodated as antisites.
Still, as in this situation the thermal formation of V$_\mathrm{Mn}$ is very inexpensive, a copious equilibrium concentration of V$_\text{Mn}$ is expected at the elevated temperatures during growth as illustrated in Fig.~\ref{figure3}(b). Diffusion measurements on MnSi have not been reported yet, but data on the isostructural FeSi suggest that diffusion in B20 compounds is extremely slow~\cite{1999:Salamon:PhilosMagA}. Hence, large thermal vacancy concentrations will be frozen in during cooling already at high temperatures.

The results of the positron annihilation experiments in combination with the calculated formation energies of point defects allow us to interpret the RRR shown in Fig.~\ref{figure1}(b). For starting compositions of $x \leq 0.01$, Si antisites (Si$_\text{Mn}$) and frozen-in thermal vacancies on the Mn sublattice (V$_{\mathrm{Mn}}$) are the dominant defect types leading to a low RRR. Around $x \approx 0.015$, at the maximum of the RRR, a pronounced drop of $c_{\mathrm{VMn}}$ is detected which indicates a transition from Mn-deficient to stoichiometric crystals with a minimal total concentration of defects. The shift with respect to $x = 0$ is attributed to the evaporation of Mn during growth. For $x \geq 0.02$, $c_{\mathrm{VMn}}$ further decreases as observed by CDBS and PALS. Inferred from the effective formation energies, the low RRR in this Mn-rich regime is mainly attributed to Mn antisites on the Si sublattice (Mn$_{\mathrm{Si}}$). It is noteworthy that contributions from vacancies on the Si sublattice (V$_{\mathrm{Si}}$) are neither detected experimentally nor expected from calculations. Contributions to the RRR arising from scattering on impurity atoms are assumed to be similar for all samples studied.

\begin{figure}[b!]
\includegraphics[width=0.85\textwidth]{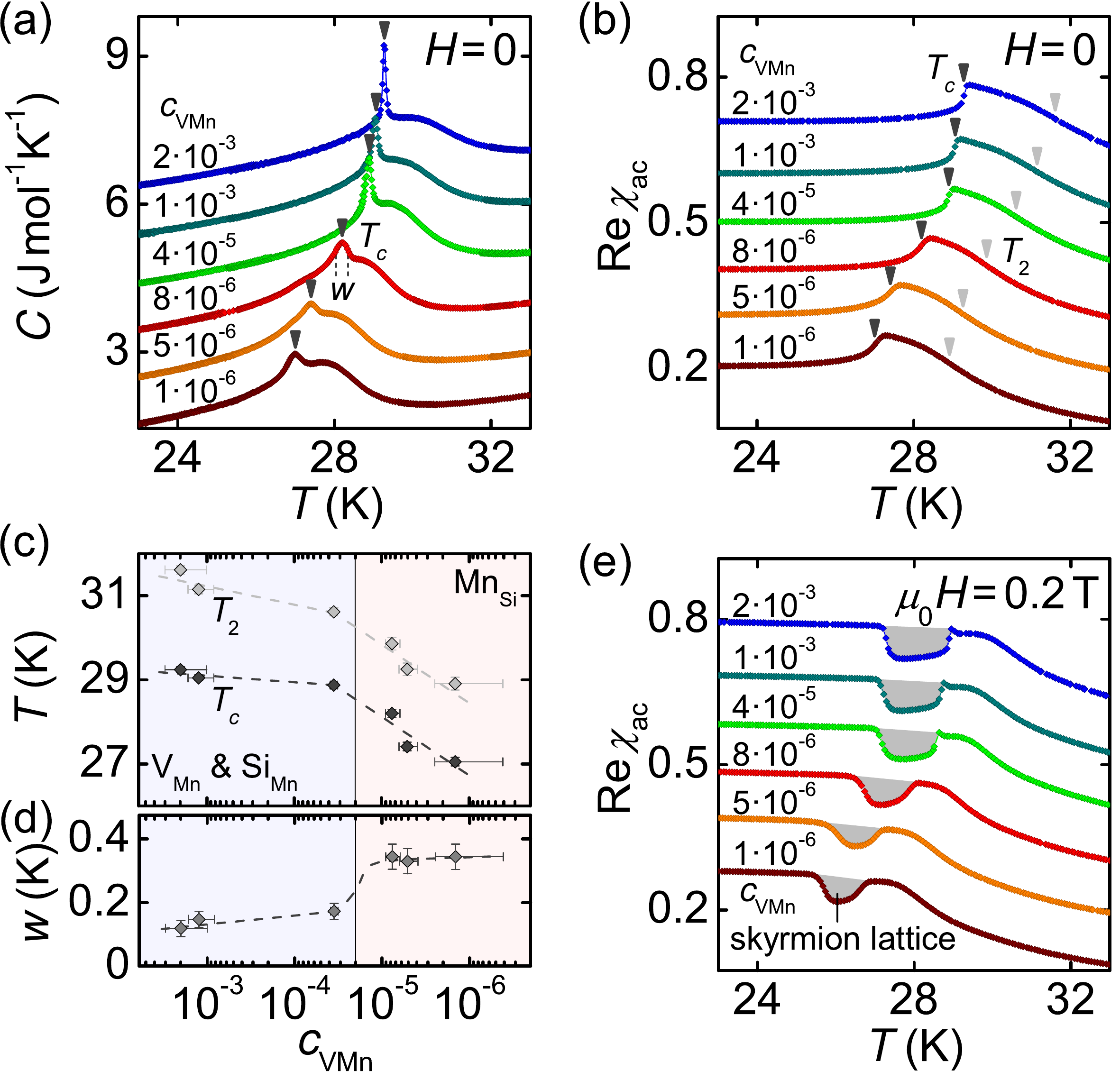}
\caption{(Color online) Magnetic properties of MnSi with different defect concentrations as determined by positron spectroscopy. \mbox{(a),(b)}~Temperature dependence of the specific heat, $C$, and the real part of the ac susceptibility, $\mathrm{Re}\,\chi_{\mathrm{ac}}$, for samples with different concentrations of Mn vacancies, $c_{\mathrm{VMn}}$. (c)~Transition temperatures $T_{c}$ and $T_{2}$ as function of $c_{\mathrm{VMn}}$. We distinguish two regimes with different dominant defect species. (d)~Temperature width, $w$, of the transition at $T_{c}$ as function of $c_{\mathrm{VMn}}$. (e)~Susceptibility in finite fields applied along $\langle100\rangle$ with a minimum attributed to a skyrmion lattice state (gray shading). Data have been offset for clarity.\label{figure4}}
\end{figure}

Figure~\ref{figure4} finally addresses the magnetic properties of different samples at temperatures around the helimagnetic phase transition. The specific heat as a function of temperature shown in Fig.~\ref{figure4}(a) is qualitatively very similar for all samples. With decreasing temperature a broad maximum is associated with the fluctuation-disordered (FD) regime, while a sharp peak marks the first-order transition at $T_{c}$. These findings are corroborated by the magnetic ac susceptibility, see Fig.~\ref{figure4}(b), where a point of inflection at $T_{2}$ defines the crossover from the paramagnet at high temperatures to the FD regime~\cite{2010:Bauer:PhysRevB, 2013:Janoschek:PhysRevB, 2013:Bauer:PhysRevLett}.

As shown in Fig.~\ref{figure4}(c), with decreasing concentration of Mn vacancies, $c_{\mathrm{VMn}}$, both $T_{c}$ and $T_{2}$ monotonically decrease. The temperature range of the FD regime, $T_{2} - T_{c}$, remains essentially unchanged. In addition, the transition at $T_{c}$ broadens as indicated by the full temperature width at half maximum, $w$, of the specific heat anomaly depicted in Fig.~\ref{figure4}(d). A prominent change of slope around $c_{\mathrm{VMn}} = 2\cdot10^{-5}$ divides our samples in two groups with either high or low concentration of Mn vacancies. As established above, Si-rich samples are dominated by Mn vacancies and Si antisites (high $c_{\mathrm{VMn}}$, blue shading) while Mn-rich specimens are dominated by Mn antisites (low $c_{\mathrm{VMn}}$, red shading). Taken together, our findings imply that in particular Mn antisite disorder leads to a suppression of the transition temperatures accompanied by a smearing of the first-order transition at $T_{c}$. Still, as depicted in Fig.~\ref{figure4}(e), all samples investigated show a plateau of reduced susceptibility just below $T_{c}$ in an applied field of 0.2\,T. This signature is characteristic of the skyrmion lattice state.

The shift of $T_{c}$ and $T_{2}$ suggests that Mn antisites directly modify the electronic structure and in turn potentially all magnetic interactions.
Apparently, ferromagnetic ordering is weakened by this kind of defects.
Understanding the microscopic mechanisms, however, will require ab-initio calculations (see e.g. Ref.~\cite{2014:Franz:PhysRevLett}).
We finally note that in MnSi defect concentrations of $10^{-3}$ and $10^{-6}$ translate to mean defect distances of 5 and 50 lattice constants, respectively, as compared to the helix wavelength of $180\,\mathrm{\AA}$ corresponding to 40 lattice constants. 
Bearing in mind the very large magnetic correlation length of ${\sim}10^{4}\,\mathrm{\AA}$ in bulk MnSi, defect-related pinning of magnetic textures is expected mainly in form of (weak) collective pinning~\cite{2011:Adams:PhysRevLett}.

\section*{Conclusion}

We have combined measurements of the magnetic bulk properties, RRR, positron annihilation spectroscopy, and calculations of effective defect formation energies in order to identify the species of point defects and their influence on magnetic and transport properties in a series of single crystals of MnSi. 
Using this approach we gained a detailed picture of characteristic lattice defects. 
Beside antisites on both sublattices, vacancies on the Mn sublattice are immanent in the system, where the dominant type of defects can be tuned by the initial Mn content.
The magnetic properties of MnSi are qualitatively extremely robust to the defect concentration including the formation of the skyrmion lattice state. However, Mn antisites shift the transition temperatures to lower values and broaden the first-order transitions. Similar consequences may also be expected for the emergent phenomena of the skyrmion lattice.
These new findings can serve as benchmark for microscopic theories on the complex magnetic behavior of MnSi.
Hence, albeit the RRR may essentially reflect the absolute density of defects, the knowledge on the type of defects is key when characterizing single crystals of MnSi.

\section*{Methods}

\subsection*{Measured Parameters}

\begin{table}[h!]
\centering
\begin{tabular}{|c|c|c|c|c|c|c|c|c|c|}
\hline
$x$ & \multicolumn{2}{c|}{RRR} & \multicolumn{2}{c|}{PAS Observables} & \multicolumn{2}{c|}{$c_{\mathrm{VMn}}$} & \multicolumn{3}{c|}{Magnetic Transition}\\
\hline
& start & end & $\eta_{\mathrm{VMn}}$ & $I_{\mathrm{VMn}}$ & discs & cylinders & $T_{\mathrm{c}}(\mathrm{K})$ & $T_{2}(\mathrm{K})$ & $w(\mathrm{K})$\\
\hline
-0.01 & 42.5 & 28.1 & 1.00 & 1.00 & $>2\cdot 10^{-3}$ & $>2\cdot 10^{-3}$ & 29.2 & 31.6 & 0.12\\
\hline
0.00 & 69.6 & 54.8 & 0.99 & 1.00 &  $1.2\cdot 10^{-3}$ & $>2\cdot 10^{-3}$ & 29.0 & 31.1 & 0.15 \\
\hline
0.01 & 109 & 135 & 0.66 & 0.80 & $3.5\cdot 10^{-5}$ & $3.3\cdot 10^{-5}$ &  28.9 & 30.6 & 0.17 \\
\hline
0.02 & 183 & 56.7 & 0.22 & 0.62 & $5.1\cdot 10^{-6}$ & $1.0\cdot 10^{-5}$ & 27.4 & 29.3 & 0.33 \\
\hline
0.03 & 94.7 & 52.7 & 0.30 & 0.68 & $7.6\cdot 10^{-6}$ & $1.5\cdot 10^{-5}$ & 28.2 & 29.9 & 0.34 \\
\hline
0.04 & 43.5 & 41.9 & 0.074 & 0.044 & $1.4\cdot 10^{-6}$ & $6.3\cdot 10^{-7}$ & 27.1 & 28.9 & 0.34\\
\hline
\end{tabular}
\caption{\label{tab:values} Physical parameters of the investigated MnSi single crystals: Initial Mn excess $x$ of sample preparation, RRR at start and end of crystals, PAS observables $\eta_{\mathrm{VMn}}$ and $I_{\mathrm{VMn}}$ determined from CDBS and PALS, respectively, (with values of 1.00 displaying saturation trapping of positrons in V$_{\mathrm{Mn}}$), vacancy concentrations $c_{\mathrm{VMn}}$ in discs and cylinders, transition temperatures $T_{c}$ and $T_{2}$ as well as the temperature width $w$ of $T_{c}$.}
\end{table}

All measured quantities are summarized in Tab.~\ref{tab:values}. The principles of the applied experimental techniques and the accompanying theoretical calculations are explained in the following.
 
\subsection*{Crystal Preparation and Characterization}
All crystal growth was carried out after evacuating the furnaces to about $10^{-7}$\,mbar and filling them with 1.5\,bar of 6N Ar treated with a point-of-use gas purifier~\cite{SAESgetter}. First, high-purity elements (precast 4N Mn and 6N Si) were alloyed in an inductively heated furnace and cooled down in less than 5\,min. In order to ensure compositional homogeneity, the resulting ingots were flipped and remolten three times before being cast to polycrystalline feed rods. Two rods of identical starting composition were optically float-zoned at a rate of 2\,mm/h or 5\,mm/h, respectively, while feed and seed rod were counter-rotating at 6\,rpm. A necking during the first millimeters of growth promoted grain selection. Despite the argon atmosphere, the high vapor pressure of Mn leads to small losses. 
All growth attempts produced single crystals of 6\,mm diameter and 10--30\,mm length with no preferred direction of growth as determined by X-ray Laue diffraction. From the start and the end of the single crystals, see Fig.~\ref{figure1}(a), we cut discs of 1\,mm thickness perpendicular to $\langle110\rangle$ using a wire saw. From each disc we prepared two platelets of ${\sim}5\times1\times0.2\,\mathrm{mm}^{3}$ with their long edge along $\langle100\rangle$. 

The RRR of these samples was measured in a 4-terminal configuration using a bespoke dipstick in a liquid helium dewar and a standard lock-in technique. Specific heat, $C$, and magnetic ac susceptibility, $\mathrm{Re}\,\chi_{\mathrm{ac}}$, were measured in a QD-PPMS on cubes of 1\,mm edge length prepared from the start discs. For the specific heat we used a quasi-adiabatic heat pulse method where pulses had a size of 30\% of the sample temperature~\cite{2013:Bauer:PhysRevLett}. The susceptibility was measured at an excitation frequency of 911\,Hz with an amplitude of 1\,mT.

\subsection*{Positron Annihilation Spectroscopy}
In general, open-volume defects such as vacancies can trap positrons prior to annihilation leading to longer lifetimes than in the bulk due to the locally reduced electron density. PALS allows to detect the resulting annihilation rates and attribute them to defect-free bulk or vacancies~\cite{2000:Coleman:Book}. In CDBS~\cite{1996:AsokaKumar:PhysRevLett,1998:Mijnarends:JPhysCondensMat} measuring the energy of both annihilation $\gamma$-quanta yields the Doppler shift $\Delta E = \frac{1}{2} p_{\mathrm{L}}c$ caused by the longitudinal momentum component, $p_{\mathrm{L}}$, of the annihilating pair. In vacancies $\Delta E$ is smaller than in bulk due to the lower overlap of the localized positron wave function with high-momentum core electrons. Furthermore, due to the intrinsically low background, CDBS allows to examine the chemical surrounding of vacancies~\cite{1996:AsokaKumar:PhysRevLett}.

CDBS was performed at the high-intensity positron beamline NEPOMUC at MLZ~\cite{2012:Hugenschmidt:NewJPhys, 2013:Reiner:JPhysConfSer} with an incident positron energy of 25\,keV corresponding to a mean implantation depth of 1.3\,$\mu$m in MnSi. Beforehand, we confirmed that the bulk of the samples was probed by variation of the beam energy. Complementary PALS was carried out at the GIPS facility by generating positrons in the bulk of the cylindrical samples from a high-energy pulsed $\gamma$-beam~\cite{2011:Butterling:NuclInstrumMethB}. The lifetime spectra were detected in a coincidence setup and analyzed by least-square fits.

In addition, we calculated CDB spectra and positron lifetimes in MnSi for annihilation in bulk as well as in vacancies V$_{\mathrm{Mn}}$ and V$_{\mathrm{Si}}$ on the Mn and Si sublattices. We used the MIKA Doppler program~\cite{2006:Torsti:PhysStatusSolidiB}, which computes the positron wave function with a two-component density functional theory in the limit of a vanishing positron density~\cite{1994:Puska:RevModPhys} and describes the electron density based on an atomic superposition method~\cite{1983:Puska:JPhysFMetPhys}. After convolution of the calculated spectra with the experimental resolution, the bulk CDB ratio curves $R_{\text{b-VMn}}$ and $R_{\text{b-VSi}}$ were evaluated using spectra for V$_{\mathrm{Mn}}$ and V$_{\mathrm{Si}}$ as reference.

\subsection*{Ab-initio Calculations on Point Defects}
Point defect energetics were calculated via density-functional theory (PBE-generalized gradient approximation~\cite{1996:Perdew:PhysRevLett}) as implemented in the ABINIT code using the projector-augmented wave framework~\cite{2009:Gonze:ComputPhysCommun}. We used a plane-wave cut-off of 30\,Ha and a shifted fcc $k$-point grid corresponding to 32 points in the Brillouin zone of the ideal B20 simple cubic unit cell. For the defect-free system, the lattice constant yielded $a = 4.499\,\textrm{\AA}$, compared to the room temperature experimental value of $4.560\,\textrm{\AA}$ \cite{Sti08}. For the point defect calculations, a bcc arrangement with four eight-atom B20 cells per supercell was chosen, with one of these 32 atoms either removed or substituted by the other constituent element. Atomic positions were relaxed under fixed supercell dimensions.
As the relevant temperature range for point defect creation and removal is far above the magnetic ordering temperature, no spin polarization was allowed.




\section*{Acknowledgements}

We thank K.\ Lochner, S.\ Mayr, C.\ Schnarr, and S.\ Zeytinoglu for assistance with the experiments and P.\ B\"{o}ni for fruitful discussions. Financial support through BMBF project no.05K13WO1, DFG TRR80, DFG FOR960, and ERC AdG (291079, TOPFIT) is gratefully acknowledged.

\section*{Author contributions statement}

M.R., A.B., and C.H. initiated the PAS experiments on defects in MnSi. M.R., A.B., and M.L. wrote the paper. A.B. grew all crystals and characterized their transport and magnetic properties. M.R. and T.G. carried out the (C)DBS experiments. M.R., W.A., M.B. and A.W. performed the PALS experiments. M.R. analyzed all data from (C)DBS and PALS experiments. P.K. was responsible for the PGAA measurements. Ab-initio calculations of the defect thermodynamics have been performed by M.L. C.P. and C.H. substantially contributed to the discussion and to the writing of the manuscript. All the authors have read the manuscript and agree with its content.

\section*{Additional information}

\textbf{Competing financial interests} The authors declare that they have no
competing financial interests.



\end{document}